%% file: nodeAware.tex
\documentclass[]{llncs}
\usepackage{graphics,graphicx}
\usepackage{amsmath}
\graphicspath{
    {.} 
    {calculations/raven_963578da_mem/analysis/memory/}
    {calculations/raven_963578da_mem/analysis/performance/}
    {calculations/stampede_email/}
}

\title{Optimizing Distributed Tensor Contractions using Node-Aware Processor Grids}
\author{Andreas Irmler \inst{1,4}\orcidID{0000-0003-0525-7772} \and
 Raghavendra Kanakagiri \inst{3} \and
  Sebastian T. Ohlmann \inst{2}      \and
 Edgar Solomonik \inst{3}        \and
 Andreas Gr\"uneis  \inst{1}
}

\institute{Institute for Theoretical Physics, TU Wien,
Vienna, Austria
\and
Max Planck Computing and Data Facility, 
Garching, Germany
\and
University of Illinois at Urbana-Champaign, USA
\and
andreas.irmler@tuwien.ac.at}

\date{today}

\usepackage{xcolor}
\usepackage[normalem]{ulem}

\newcommand{\TODO}[1]{TODO: \textcolor{red}{#1}}

\begin{document}
\maketitle
\begin{abstract}
We propose an algorithm that aims at minimizing the inter-node communication volume for distributed and memory-efficient tensor contraction schemes on modern multi-core compute nodes.
The key idea is to define processor grids that optimize intra-/inter-node communication volume in the employed contraction algorithms. We present an implementation of the proposed node-aware communication algorithm into the Cyclops Tensor Framework (CTF).
We demonstrate that this implementation achieves a significantly improved performance for matrix-matrix-multiplication and tensor-contractions on up to several hundreds modern compute nodes compared to conventional implementations without using node-aware processor grids.
Our implementation shows good performance when compared with existing state-of-the-art parallel matrix multiplication libraries (COSMA and ScaLAPACK).
In addition to the discussion of the performance for matrix-matrix-multiplication, we also investigate the performance of our node-aware communication algorithm for tensor contractions as they occur in quantum chemical coupled-cluster methods. To this end we employ a modified version of CTF in combination with a coupled-cluster code (Cc4s). Our findings show that the node-aware communication algorithm is also able to improve the performance of coupled-cluster theory calculations for real-world problems running on tens to hundreds of compute nodes.
\end{abstract}

\section{Introduction}
Matrix-matrix multiplication (MMM) is ubiquitous in the field of
scientific computing, computational physics, machine learning
and many other areas of significant technological and scientific relevance.
One important area of application of MMM in physics includes electronic structure theory, which is part of the motivation for the present work.
We note that electronic structure theory calculations often
involve operations on large matrices that need to be
distributed over many tens to hundreds of modern compute nodes in order
to satisfy memory requirements. Therefore, 
electronic structure theory calculations have evolved in parallel
to hardware improvements and newly developed efficient linear algebra libraries
over the past few decades.
In this paper, we seek to compare and improve algorithms employed in popular
MMM libraries including ScaLAPACK~\cite{ScaLAPACK}, COSMA~\cite{Kwasniewksi2019} and CTF~\cite{Solomonik2014}.
In particular, we focus on the effect of network contention
and inter-node communication within CTF.

In addition to MMM, the present work seeks to extend the presented development
to more general tensor algebraic operations.
We note that tensor algebra is yet another important technique widely-used
in electronic  structure theory, especially for highly accurate and 
computationally expensive many-electron methods.
With the development of more sophisticated distributed
tensor algebra libraries, however, their implementation becomes 
simpler and allows for efficient calculations
of increasingly large problems on modern HPC clusters.

We also demonstrate a real-world application that involves coupled-cluster theory
calculations. Coupled-cluster theory is a many-electron perturbation theory, which
is widely-used in the  field of computational chemistry and many-body physics.
The solution of the underlying set of nonlinear equations involves tensor
contractions. Already for the study of relatively few atoms,
the memory footprint of the required tensors typically exceeds even the main memory of modern
nodes. Furthermore the computational cost
required by these calculations also grows rapidly with the number of atoms.
This necessitates implementations of coupled-cluster 
methods
employing massive parallel tensor contraction libraries. 
Our node-aware CTF implementation shows speed-ups of up to 3X relative to the prior node-oblivious implementation,
for real-world coupled-cluster
theory calculations.

Overall, our paper introduces the following contributions:
\begin{itemize}
\item node-aware parallel algorithms for matrix multiplication and tensor contraction, which minimize inter-node communication volume,
\item an implementation of these algorithms as part of the Cyclops library,
\item an experimental evaluation comparing the implementation to other codes on two supercomputers and as part of a quantum chemistry method.
\end{itemize}

\section{Node-Aware Multiplication and Contraction}
Distributed-memory algorithms for matrix multiplication generally aim to
minimize communication cost (in terms of latency, i.e., the number of messages,
and bandwidth cost, i.e., the amount of data sent).  Communication cost in this
setting is often measured by the amount of matrix entries (words) sent and
received by each processor, with matching sends and receives assumed to execute
concurrently.  In the memory-constrained setting, for multiplication of
$n\times n$ matrices, Cannon's algorithm~\cite{Cannon:1969} achieves a
communication cost of $O(n^2/\sqrt{p})$ when running with $p$ processors.  This
cost is optimal according to known lower
bounds~\cite{irony2004communication}.
In practice, the SUMMA algorithm~\cite{VanDeGeijn1997,Agarwal1995} or variants
thereof are most often implemented in libraries (e.g., ScaLAPACK and CTF both
use SUMMA).  The SUMMA algorithm leverages broadcasts and reductions, which
have a slightly higher latency (require $O(\log p)$ messages) than the
point-to-point messages used in Cannon's algorithm.  However, large-message
broadcasts and reductions can be done with the same asymptotic bandwidth cost
as sends, $O(n)$ for a message of size $n$, so long as $n=\Omega(p)$~\cite{chan2007collective,thakur2003improving}.
Further, the SUMMA algorithm is easier to extend to nonsquare matrices
than Cannon's approach, and use of similar collective communication also allows
for implementation of 3D algorithms, which minimize communication cost when
additional memory is available~\cite{aggarwal1989communication,Agarwal1995,mccoll1999memory,Solomonik2011}.

On modern supercomputers and clusters, each node contains many CPUs and/or
GPUs.  Even with the use of threading, most MPI-based codes achieve highest
efficiency when executed with multiple MPI processes per node (e.g., one per
GPU or one per NUMA region).  Given the presence of multiple communicating
processes per node, the performance of collective communication operations, such
as broadcast and reduction, become dependent on the number of distinct nodes in
the subcommunicator used for the operation.  In particular, while we have
mentioned that the per-processor communication-cost is largely independent of
$p$, the communication volume (total number of words sent or received by any
processor) for a broadcast of a message of $n$ words to $p$ nodes is $n(p-1)$.
Unlike per-processor communication cost, communication volume does not directly
model runtime, but higher communication volume entails increased contention for
network and injection bandwidth.  We propose an algorithm to select an
MPI-process-to-node mapping that minimizes the communication volume for dense
matrix multiplication (and later tensor contractions) executed on any given
initial processor grid.  Similarly motivated node-aware optimizations have
previously been presented for accelerating point-to-point communication in
sparse matrix vector products~\cite{Lockhart2023,Lockhart2022,Bienz2019,Williams2007,Bienz2020}.

\subsection{Node-Aware Matrix Multiplication}

We first propose a scheme to map processes to nodes for matrix multiplication,
aiming to accelerate 2D (SUMMA) and 3D matrix multiplication algorithms used by
CTF~\cite{Solomonik2014}.  CTF generally selects a 3D processor grid $p_1\times
p_2 \times p_3$ (1D or 2D processor grids may be obtained by setting of $p_1$,
$p_2$, and $p_3$ to 1) at runtime so as to minimize cost (based on not just
communication, but a more detailed performance model that includes predicted
cost of local work and redistribution).  All communication within the matrix
multiplication algorithm is performed by concurrent broadcasts and reductions
among fibers of the 3D processor grid (e.g., $p_1p_2$ concurrent broadcasts
with $p_3$ processors involved in each).  Once a processor grid mapping is
selected, the counts of words communicated along each fiber $W_1$, $W_2$, and
$W_3$ are known.  When executing with $m$ processors per node, we consider the
best choice of an $m_1\times m_2\times m_3$ intra-node processor grid with
$m_1m_2m_3=m$ and $p_i\equiv 0 \bmod m_i$, for all $i$.  The $p/m$ nodes are
then arranged in a 3D processor grid of dimensions $p_1/m_1 \times p_2/m_2
\times p_3/m_3$, so that each original fiber of size $p_i$ stretches across
$p_i/m_i$ physical nodes.  We choose the intra-node
processor grid, so as to minimize the communication volume, \[V =
W_1(p_1/m_1-1) + W_2(p_2/m_2-1) + W_3(p_3/m_3 -1).\] Once the mapping is
chosen, we redistribute the tensor data, which can be done with a single round
of concurrent point-to-point messages (each processor sends all of its matrix
data to a single other processor in the new mapping).

\subsection{Node-Aware Tensor Contractions}

CTF leverages nested SUMMA, in combination with 1D replication/reduction, to
generalize 2D and 3D algorithms for matrix multiplication to tensor
contraction~\cite{Solomonik2014}.  Processor grids $p_1\times \cdots \times p_d$ with $d>3$
are used to accommodate nested SUMMA and to support symmetric-packed tensor
formats efficiently (only unique entries of a symmetric tensor are stored by
CTF, e.g., only the lower triangular part of a symmetric matrix).  Each of
these matrix multiplication variants results in some amount of words broadcast
or reduced along each processor grid fiber, say $W_i$ along fiber $p_i$.  Our
node-aware mapping algorithm proceeds analogously to the matrix multiplication
case.  We select the best choice of $m_1\times \cdots \times m_d$ intra-node
processor grid and combine it with a $p_1/m_1\times \cdots \times p_d/m_d$
inter-node processor grid, so that the $i$th fiber of the grid spans $p_i/m_i$
distinct nodes.  Again, we select the processor grid to minimize the
communication volume, \(V=\sum_{i=1}^d W_i(p_i/m_i-1).\)

To find the optimal cost configuration, we use exhaustive search.  We enumerate
all distinct factorizations of $m=m_1\cdots m_d$ such that $p_i \equiv 0 \bmod
m_i$.  Provided a model of the affect of communication volume on runtime, this
search could be done together with the selection of the best processor grid
$p_1\times \cdots \times p_d$ and the tensor mapping.  However, searching this
larger space of mappings would be computationally expensive.  Specifically, if
$K$ processor grids and distinct mappings are considered and $L$ virtual
processor grids are considered, the combined search space is of size $O(KL)$
instead of $O(K+L)$.

\section{Evaluation Methodology}
\vspace{-0.1in}
\subsection{Hardware and Software Platform}
Results are collected on the CPU partition of the Raven supercomputer at the Max Planck Computing and Data Facility. It consists
of 1592 compute nodes; each node has an Intel Xeon IceLake-SP Platinum 8360Y processor with
72 cores and 256 GB RAM per node. 
As interconnect, it uses Mellanox HDR
InfiniBand network (100 Gbit/s) with a pruned fat-tree topology and
non-blocking islands of 720 nodes; all jobs run inside one island. 
The theoretical peaks of floating point operations and memory bandwidth are 5.53
TFLOP/s and 320 GB/s per node, respectively.
To demonstrate the robustness of our approach, we also collect results (for a subset of the experiments) on the Stampede2 supercomputer. Each node has an Intel Xeon Phi 7250 CPU with 68 cores, 96GB of DDR4 RAM
Note that Stampede2 has a distinct configuration when compared to that of Raven.

We evaluate our node-aware version of CTF (CTF-na) by comparing against the default CTF (CTF-def)~\cite{Solomonik2014}, ScaLAPACK~\cite{ScaLAPACK}, and COSMA \cite{Kwasniewksi2019}.
We use ScaLAPACK as provided by Intel MKL (version 2022.0).  All codes were compiled using the Intel compiler (version
2021.5) and Intel MPI (version 2021.5).
In all our calculations, we use 
one core per MPI rank.
All codes would in principle
allow a hybrid OpenMP/MPI approach. 
In COSMA, the authors note that their strategy performs best with one core per rank \cite{Kwasniewksi2019}.
Our tests show that CTF performs equally good with one to four cores per rank.


COSMA allows communication-computation overlap. Our tests show that for the
chosen matrix dimensions the results with and without overlap strategy are
very similar. The differences are at most in the order of 5\%. For a more expressive comparison against CTF and ScaLAPACK, both of which do not offer overlapping strategies, we do not use computation-communication overlap
in any of our COSMA calculations.
Furthermore, it is possible to adjust the used memory in a COSMA calculation.
More memory possibly allows to employ a more efficient parallelization
strategies, viz. a higher performance. In this work, we use two values for the
allowed memory. The lower limit is chosen to be  $2.5$-times the size storing
the three matrices.  The upper limit is chosen to be such that the full memory
of the machine can be utilized.  In the following, we will label these schemes
as COSMA-lim and COSMA-unl

\subsection{Matrix-Multiplication Benchmarks}
In this section, we present details about the dataset used for our main results, which is collected from the Raven cluster.
We investigate four cases of products of an $m\times k$ matrix with a $k\times n$ matrix, namely, square ($m=n=k$),
large $K$ ($m=n \ll k$), large $M$ ($m \gg n = k$), and small $K$ ($m=n \gg k$).  
The ratio between small and large edge is chosen to be
$10$ for all systems. 
We exploit results for different number of nodes ranging
from 1 to 288 nodes
using all node numbers which fulfill the
following equation $n = j \cdot 2^i \text{, with: } j \in [1,3,9]$. 
We consider both strong and weak scaling in our experiments. For strong scaling, we choose the dataset size such that it is just large enough to be stored (and contracted) on a single node which is approximately 150~GB.
For weak scaling, we use two different matrix dimensions (sizes), 
such that the matrices occupy either 0.5\% or 5\% of the overall system memory. 
In subsequent sections, we denote the strong scaling results as ``strong", while the weak scaling datasets are referred to as ``weak18" and ``weak180", corresponding to the 0.5\% and 5\% memory occupation, respectively.

\subsection{Experimental Methodology}
\label{sec:exp_methodology}
For each combination of parameters considered, we perform five contractions (runs) using each of the five strategies (CTF-def, CTF-na, ScaLAPACK, COSMA-lim and COSMA-unlim) on the same node allocation (i.e. via a single job submission to the cluster).
We exclude the two slowest runs and compute the average based on the remaining three runs.
We submit each job twice in order to have two random
node allocations. 
Consequently, all presented data points are mean values
averaged from 6 calculations, each. Typically, the standard deviation of the
mean value is below 1 GFLOPS/core, so we do not include any error bars.
In the presented results node-aware redistribution is performed whenever a topology with a lower inter-node communication is found. We stress that this might not always be an optimal strategy since it does not take into account the time required for redistributing the data from the default to node-aware topology before performing the contractions. When we integrate our strategy in the coupled-cluster calculations, the performance model accounts for the redistribution time thus finding the overall optimal solution.

\section{Performance Results/Evaluation}
\vspace{-0.1in}
\subsection{Memory Footprint}
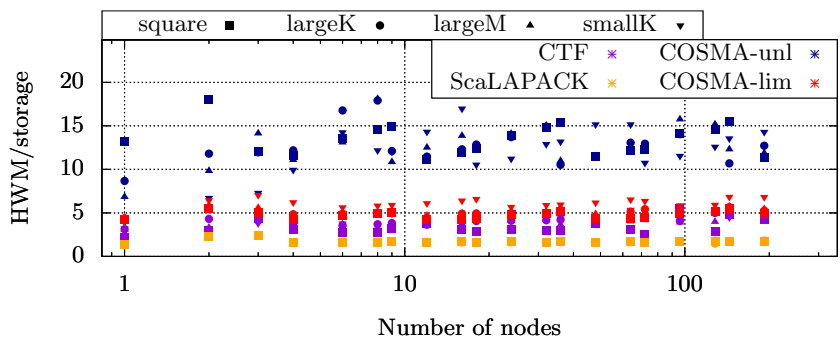
\begin{figure}
\centering
\input{result_memana}
\caption{\label{fig:memana}
Ratio of computation's HWM over storage size for different node counts. 
}
\vspace{-0.2in}
\end{figure}

Prior to comparing the performance of the various implementations, we analyze their memory requirements. 
We refer to the maximum memory consumption by the implementation (when executing the contraction) as high-water mark (HWM).
In Figure~\ref{fig:memana}, we present HWM 
for weak180 calculations for all considered matrices, representing the maximum memory consumption. If we exclude the results for one to three nodes, we observe that ScaLAPACK maintains a nearly constant ratio of HWM over storage size, averaging around 1.66. This is true for all type of matrix contractions. For CTF, the
ratio is between 2.5 and 5, depending on the number of nodes and the
contraction type. The additional memory overhead
compared to ScaLAPACK is explained by the extra redistribution buffers and the
2.5D algorithm. 
COSMA-lim shows a very similar HWM as CTF with values between
4 and 7 for the ratio HWM over storage size. This enables a fair comparison
between CTF and COSMA in the case of similar memory constraints. 
Disregarding a
handful of outliers COSMA-unl shows a ratio between 10 and 18. We recall that for
these calculations the storage size is 5\% of the main memory, implying that
the COSMA calculation utilizes a large fraction of the total main memory. We
note that CTF-na has the same memory footprint as CTF-def.

\begin{figure*}
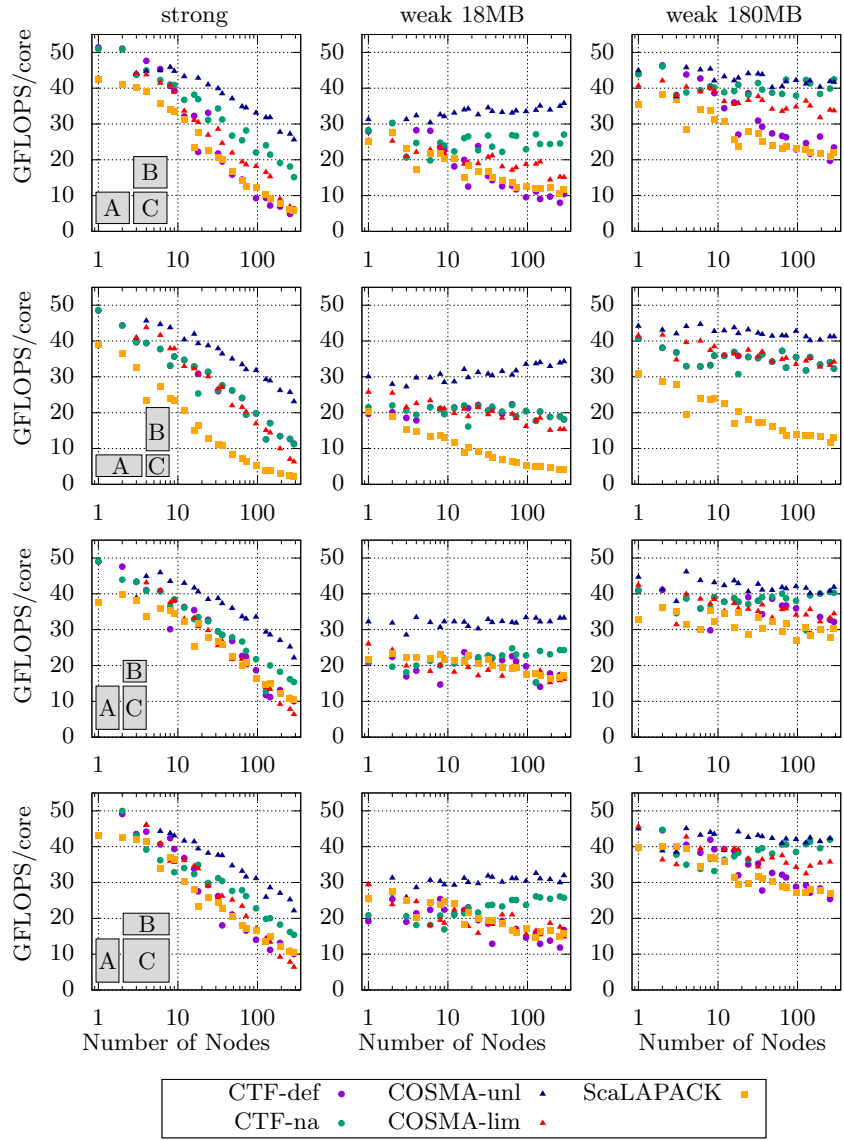

\centering
\include{result_matrix}
\vspace{-0.7in}
\caption{\label{fig:matrix}
Performance in GLFOPS/core on RAVEN.
From top to bottom, rows shows results for different matrix-matrix products: square, large $K$ , large $M$, and small $K$.  
}
\end{figure*}

\subsection{Matrix multiplication}
In the section, we present one of the primary results of this work.
Figure~\ref{fig:matrix} shows the performance results for all the matrix sizes and implementations considered (see Section \ref{sec:exp_methodology}).
%
We first note that for all contraction types and all
scaling scenarios, COSMA-unl achieves the best performance.
The improvement over the second best method is very pronounced for
situations where the operations per core are low, i.e. large node numbers in
strong scaling scenario and for the weak scaling scenario with 18~MB. For the
weak scaling scenario with 180~MB, the improvements are much smaller.
We note, however, that the memory footprint of COSMA-unl is relatively high in all calculations as depicted in the previous sections.
The goal of the present work is to advance memory efficient
tensor contraction algorithms with high scalability on multi-core nodes. \\
\underline{\textbf{Square}}: 
The performance for the square contractions is shown in the top panels of
Figure~\ref{fig:matrix}. When employing more than ten nodes, CTF-na shows the second
best performance followed by COSMA-lim, ScaLAPACK, and CTF-def exhibit a very
similar performance.
For small node
numbers the same trend is generally present, however, the results are way more
noisier here.
We note that CTF-na is particularly efficient for the large memory
weak scaling scenario (180MB). Large square MMM present one of the best
application regimes of CTF-na compared to the even more efficient but memory intense
COSMA-unl implementation.

The node-aware algorithm significantly improves the results compared to the
results with CTF in default topology. For more than 50 nodes the performance
improves by a factor of 1.5-5.5X in weak and strong scaling scenarios.
Further, CTF-na outperforms COSMA-lim when using more than ten nodes.\\
\underline{\textbf{Large $K$}}: 
For the large $K$ contraction
(second row of Figure~\ref{fig:matrix})
COSMA-unl achieves the best performance and
ScaLAPACK the worst performance. 
Here COSMA-lim, CTF-def, and CTF-na show very similar results.
There are two reasons why CTF outperforms ScaLAPACK significantly for this contraction
type.  Firstly, within this contraction, CTF communicates the matrix C as it is the
smallest occurring matrix. Secondly, CTF employs the SUMMA 2.5D algorithm. 
In this case, the node-aware topology does not lead to any
further improvements of the CTF-def algorithm.
The reason for this is that the default processor grid for these contractions already achieves low inter-node communication volume.\\
\underline{\textbf{Large $M$}}:
The performance for the large M contractions is shown in the third row of
Figure~\ref{fig:matrix}.
Once more, COMSA-unl exhibits throughout the best performance for all
calculations. However, all four other implementations show similar
results. CTF-na shows an improvement over CTF-def only on some node counts. \\
\underline{\textbf{Small $K$}}: 
The performance for the small $K$ contractions is shown in the bottom row of
Figure~\ref{fig:matrix}.
The small $K$ results are similar to the results of the square contraction. The
node-aware topology outperforms the default calculation especially for large
number of nodes. COSMA-unl outperforms all other implementations in the
strong scaling regime, as well as for the weak scaling scenario with 18~MB.
However, for the 180~MB scenario and more than 50 nodes CTF-na achieves very
similar results as COSMA-unl
\begin{figure*}
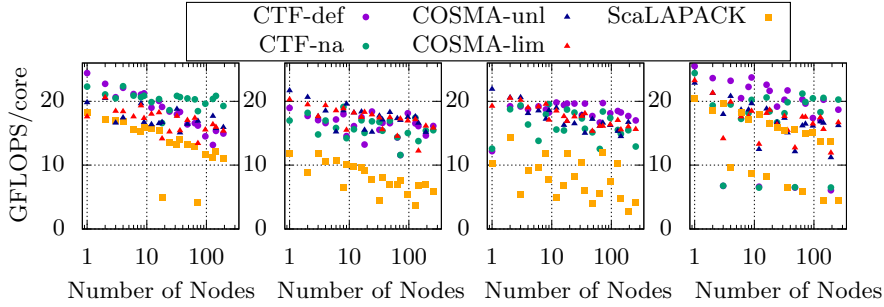

\centering
\include{result_stampede}
\caption{\label{fig:matrix_stampede}
Achieved performance on Stampede2 measured in GLFOPS/core.
From left to right each column shows results for a different matrix
contractions: square, large $K$ , large $M$, and small $K$.
The results use matrix sizes of 80~MB per processor.
}
\vspace{-0.2in}
\end{figure*}
\\
\underline{\textbf{Weak scaling performance on Stampede2}}: 
In addition to the results obtained on Raven, we have also investigated
the performance of the different MMM libraries on Stampede2.
The Stampede2 compute nodes are equipped with significantly less main memory
than Raven nodes, making it more difficult to perform calculations with implementations
that exhibit a large memory footprint such as COSMA.
Figure~\ref{fig:matrix_stampede} depicts performance results for weak scaling. \\
\textbf{Square and Small $K$}: The results obtained are similar to those
obtained for Raven. ScaLAPACK exhibits the worst performance.
CTF-na improves significantly over CTF-def
for large numbers of nodes.
COSMA-unl and COSMA-lim perform slightly worse than CTF-na
for large numbers of nodes.
While COSMA-unl exceeded available memory in some cases, it outperforms COSMA-lim in most cases.
We also note that some node counts
exhibit a much poorer performance for all employed
libraries. A more careful analysis of these outliers
reveals that this reduction is not caused by increased
communication volume, but by performance drops in the
GEMMs. \\
\textbf{Large $K$}:
Here CTF-na shows no improvement
over CTF-def and ScaLAPACK yields the lowest performance. Interestingly the
performance of COSMA is very similar, whereas on the Raven system 
COSMA clearly outperformed CTF.
\\
\textbf{Large $M$}:
The large $M$ contractions on Stampede2 show similar patterns to Raven.
CTF-na is not improving over CTF-def for large node counts due to efficiency of the default mapping of CTF-def. \\
\newline
\textbf{Summary}:

We now summarize the most important findings for the performance analysis.
Table~\ref{tab:perfMatrix} lists mean values of the achieved speedups for CTF-na
compared to the four other implementations.  Averaged results are provided for
calculations employing less and more than 50 nodes, respectively.  The values
are smaller or equal than 1 only for the case of COSMA-unl, indicating that
COSMA-unl achieves in all scenarios the best performance compared to the other
methods at the price of a larger memory footprint.  All other reported values
are equal to 1.0 or larger than 1.0, implying that CTF-na achieves the same or
better performance than CTF-def, COSMA-lim and ScaLAPACK most cases.  Compared
to ScaLAPACK the speedup is on average between 1.4 and 4.1 for square, large
$K$ and small $K$ for all scenarios when using more than 50 nodes.  The speedup
compared to ScaLAPACK is only about 1.3 in the case of large $M$.  Compared to
COSMA-lim, CTF-na achieves on average a speedup between 1.2 and 1.5 for more
than 50 nodes in the cases of square and small $K$ MMMs. This speedup reduces
to about 1.0 to 1.2 in the cases of large $K$ and large $M$.  Similarly,
compared to CTF-def, which disregards node-awareness, we see significant
speedups for square and small $K$ contractions. Whereas only minor improvements
for large $K$ and large $M$ contractions.

\begin{table}[t]
\centering
\label{tab:perfMatrix}
\caption{
Measured speedup of CTF-na on Raven. compared to the other algorithms. For the different
scenarios and contraction types. Averaged values are provided for less and more
than 50 nodes, respectively.
}
\begin{tabular}{ |l | l | cc |cc | cc |}
\hline\hline
         & & \multicolumn{2}{c|}{strong}  & \multicolumn{2}{c|}{weak 18} & \multicolumn{2}{c|}{weak 180} \\
         & &$N<50$ &$N>50$&$N<50$& $N>50$ & $N<50$ &$N>50$ \\
\hline
square & CTF-def  & 1.2&2.6 & 1.3&2.5 & 1.1&1.7 \\
& ScaLAPACK & 1.3&2.3 & 1.3&2.1 & 1.4&1.8 \\
& COSMA-unl& 0.9&0.7 & 0.8&0.8 & 0.9&1.0 \\
& COSMA-lim& 1.1&1.8 & 1.2&1.5 & 1.1&1.2 \\
\hline
large $K$ & CTF-def  & 1.0&1.0 & 1.0&1.0 & 1.0&1.0 \\
& ScaLAPACK & 1.8&4.1 & 1.9&3.9 & 1.7&2.5 \\
& COSMA-unl& 0.8&0.6 & 0.7&0.6 & 0.8&0.8 \\
& COSMA-lim& 1.0&1.3 & 1.0&1.2 & 0.9&1.0 \\
\hline
large $M$ & CTF-def  & 1.0&1.3 & 1.0&1.3 & 1.0&1.2 \\
& ScaLAPACK & 1.2&1.3 & 1.0&1.3 & 1.2&1.3 \\
& COSMA-unl& 0.9&0.7 & 0.7&0.7 & 0.9&0.9 \\
& COSMA-lim& 1.1&1.6 & 1.1&1.3 & 1.0&1.1 \\
\hline
small $K$ & CTF-def  & 1.1&1.6 & 1.1&1.7 & 1.0&1.4 \\
& ScaLAPACK & 1.2&1.5 & 1.0&1.6 & 1.1&1.4 \\
& COSMA-unl& 0.8&0.7 & 0.7&0.8 & 0.9&1.0 \\
& COSMA-lim& 1.0&1.7 & 1.0&1.5 & 1.0&1.2 \\
\hline
\hline
\end{tabular}
\end{table}

\if
\input{node_aware_bcast.tex}
\fi

\section{Performance of coupled-cluster calculations}
\begin{figure*}[t]
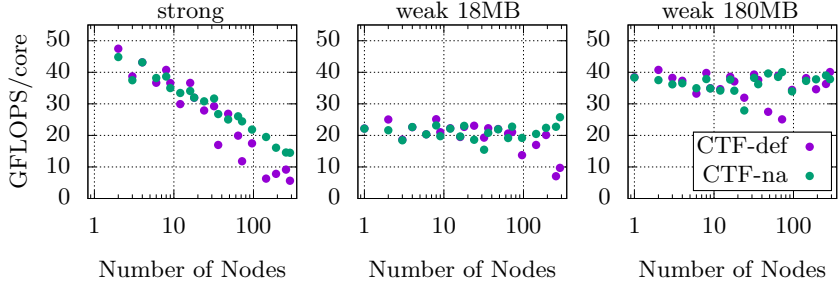

\centering
\include{result_ring}
\caption{\label{fig:ringContraction}
Results for the drCCD contraction on RAVEN.
}
\vspace{-0.2in}
\end{figure*}

We now present results for more general tensor contractions, going beyond
matrix matrix multiplications.  This section presents performance results
obtained for coupled cluster calculations as implemented in the Cc4s code
\cite{cc4s}, which employs the CTF library.  CC methods are widely
used in the field of electronic structure theory to study many-electron systems
\cite{Bartlett2007}. From a computational perspective, CC methods involve high
order tensor contractions.  The CC method, which employs single and double
particle-hole excitation operators, is called CCSD.  The computational cost of
a CCSD calculation is dominated by solving the nonlinear doubles amplitude
equations given by

\begin{equation}
D_{ij}^{ab} t_{ij}^{ab} = v_{ij}^{ab} + 2 \sum_{ck} v_{ic}^{ak} t_{kj}^{cb}
                        + \sum_{klcd} t_{il}^{ad} v_{dc}^{lk} t_{kj}^{cb}
                        + \sum_{cd} v_{cd}^{ab} t_{ij}^{cd}
                        - \sum_{ck} v_{ic}^{ak} t_{kj}^{bc} + ... \text,
\end{equation}
where the dimensions are chosen such that $\textrm{dim}(i)=\textrm{dim}(j)=\textrm{dim}(k)=\textrm{dim}(l)$
and $\textrm{dim}(a)=\textrm{dim}(b)=\textrm{dim}(c)=\textrm{dim}(d)$.
The amplitude equations are solved iteratively employing a Jacobi method
such that most of the computational cost originates from tensor contractions
as defined by terms on the right-hand-side of the above equation.
CCSD exhibits a memory footprint and computational cost that scales as
$\mathcal{O}(N^4)$ and $\mathcal{O}(N^6)$, respectively.
$N$ is proportional to the number of electrons in the system.
The dimension of the indices $i,j,k,...$ and  $a,b,c,...$  corresponds to the
number of occupied orbitals and the number of virtual orbitals, respectively.
In a typical calculation, the number of virtual orbitals is
10-30 times larger than the number of occupied orbitals.
As a result the so-called particle-particle-ladder (ppl) term,
\(
r_{ij}^{ab} = \sum_{cd} v_{cd}^{ab} t_{ij}^{cd},
\)
is treated in a special way to avoid storing the tensor $v_{cd}^{ab}$ in main memory.  This is achieved by
computing slices of $v_{cd}^{ab}$ on-the-fly and
contracting them consecutively.

In addition to the CCSD method, we also investigate the performance of drCCD,
which is a popular approximation to the CCSD method. The drCCD method only
includes so-called ring diagrams in the amplitude equations, corresponding to
terms given by 
\(
r_{ij}^{ab} = \sum_{ck} v_{ic}^{ak} t_{kj}^{cb}.
\label{eq:contr_ring}
\)

\begin{table}
\centering
\label{tab:cc4s}
\caption{
Performance given in GFLOPS/core from different coupled-cluster calculations for three different node counts.
}
\begin{tabular}{l||cc|cc|cc}
  & \multicolumn{2}{c}{32 nodes}     &\multicolumn{2}{c}{72 nodes}  & \multicolumn{2}{c}{128 nodes} \\
  & \multicolumn{2}{c}{dim($i$)=116}   &\multicolumn{2}{c}{dim($i$)=142 }  & \multicolumn{2}{c}{dim($i$)=164} \\
    & \multicolumn{2}{c}{dim($a$)=1161 } &\multicolumn{2}{c}{dim($a$)=1422 }  & \multicolumn{2}{c}{dim($a$)=1642} \\
\hline
Method        & default & node-aware & default & node-aware & default & node-aware \\
\hline
CCSD          & 19.4    & 21.0       &  20.5   & 24.0     &  25.0   &  25.0      \\
CCSD (no ppl) & 26.6    & 32.6       &  25.1   & 37.0     &  37.3   &  37.5      \\
drCCD         & 37.4    & 37.9       &  24.7   & 39.0     &  38.6   &  38.5      \\
\end{tabular}
\end{table}

We now seek to discuss the performance of the following types of CC
calculations: drCCD, CCSD and CCSD excluding the ppl-term. Performance results
have been obtained using the default CTF version and the node-aware CTF
version.
Figure~\ref{fig:ringContraction} depicts the performance of drCCD calculations in
strong and weak scaling scenarios. Our findings show that the improvements are
not as pronounced as for the case of MMMs. Only for the strong scaling case we
observe significant improvements when comparing CTF-na to CTF-def for about 100
nodes.  For the weak scaling case with smaller problem sizes, CTF-na achieves
no significant improvements compared to CTF-def. For weak
scaling case, CTF-na improves the performance of CTF-def for only a few  cases.
An analysis of these cases (such as, for example
48 and 72 nodes) reveals that they
correspond to core counts for which the default topology leads to a
significantly increased inter-node communication volume compared to the
node-aware topology. However, CTF-na is able to achieve excellent and stable performance
for all node numbers. For practical calculations that cannot be
tuned to circumvent special node numbers, where CTF-def performs less efficient, CTF-na presents a valuable
improvement.  We find that higher-order tensors are more often
already distributed in an communication efficient manner using the default
topology, which is why the node-aware distribution often has a
negligible effect.
However, we find for some node counts significant improvement up to 3X.
We also explore the performance of CCSD calculations.  These
calculations are computationally even more expensive than drCCD.
Table~\ref{tab:cc4s} lists results for a selected number of nodes, including
special cases described above for drCCD.  In addition to CCSD calculations, we
also measure the performance of CCSD excluding the ppl-term (CCSD no ppl).  
The presented results imply that the evaluation of the ppl-term is performed at lower efficiency than the other tensor contractions.
The cause for the bad performance is not related to the
node-aware topology and will be explained in future work.
Consequently, we will restrict the following discussion on CCSD calculation excluding the ppl-term.

Similar to drCCD,
observe performance improvements of CCSD calculations when
using CTF-na instead of CTF-def  only for some node numbers.
There are cases where a drCCD calculation is not improved by
node-awareness, whereas the CCSD calculation
improves by 10-20\%. This is because the processor grid for every single
contraction is determined on runtime and generally differs for different
contractions as they appear in the CCSD equations.

\section{Related Work}
There have been several works that derive communication-optimal algorithms for matrix-matrix multiplication \cite{carma,Kwasniewksi2019,Solomonik2011}.
CARMA~\cite{carma} has provided the first approach to minimize communication for any $M$, $N$, $K$ and any number of processors/available memory.
COSMA \cite{Kwasniewksi2019} provides a theoretically optimal distributed dense matrix-matrix algorithm as well as the current best known implementation. Similar to CTF, COSMA finds the best layout via a cost model subject to memory constraints. It  leverages RDMA, and a custom implementation of a binary tree collective. It also proposes to overlap communication with computation. 
Both CTF and COSMA rely on an analytical model and minimize the communication cost.
In this work, we explore a cost model that goes beyond what is considered in CTF and COSMA. We take into account the communication cost not just between MPI processes but also across nodes in the network.
Further, we are able to obtain nearly the same performance and in some cases better, without low-level optimizations that may be less portable.

\if
COSMA specifically employs a strategy where an optimal schedule is derived based on a red-blue pebble game. The optimal schedule considers both intra-node I/O (load and store operations to memory) and inter-node communication. Each processor is then assigned a portion of the computation such that each process gets assigned equal memory and computation.
This is in contrast to the techniques that libraries like ScaLAPACK and CTF employ, wherein, a suitable processor grid is chosen based on the matrix dimensions and then the decomposition of the computation is mapped to the grid. \\
AI: Though COSMA employs an optimal I/O schedule, this does not necessarily translate to the best performance in an actual calculation. This is because on modern hardware communication performance is not just dependent on the message size. On the contrary, there is a dependence to which rank a message is sent -- it can be on the same socket or even on a different node. Furthermore there are optimized communication schemes for different communication patterns. MPI collectives for instance are highly optimized implementations of well defined communication patterns. These are significantly faster than the equivalent naive point-to-point communication. Restricting the analysis to finding the lowest I/O schedule can not capture these intricate effects.
\\
AI:Many MMM algorithms try to efficiently utilize possible extra memory to decrease communication and thereby increase the performance. However, in a wide range of applications memory is scarce, so the question is how different algorithms perform under different memory constraints. That is why we include the memory footprint in our comparison of different algorithms
and choose our weak scaling scenario such that the storage size of the participating matrices is kept constant when probing different number of nodes. This is specifically important in coupled-cluster calculation - an application which is discussed later in this work.
\fi

In \cite{Bienz2019}, the authors propose a node-aware sparse matrix-vector multiply, where values are gathered in processes local to each node before being sent across the network, followed by a redistribution at the receiving node. This optimized point-to-point communication leads to reduction in communication cost. A similar technique is used in \cite{Lockhart2023} when using enlarged conjugate gradient methods. 

\section{Conclusion}
In this work we have presented a modification to the Cyclops Tensor
Framework that employs node-aware processor grids.
We have shown that the achieved performance improvements
due to the node-aware topology in CTF are most strongly
pronounced in the case of square and small $K$ matrix-matrix products.
In the case of large $K$ and
large $M$ matrix multiplication, the default processor grids employed by CTF are already efficient.
%
Although the memory-unlimited version COSMA achieves overall the best performance for matrix multiplication, CTF with node-awareness is competitive and often more performant when the same memory limit is imposed on COSMA.

In addition to the results for MMMs, we have also investigated
the performance of the modified version of CTF for tensor contractions
in coupled-cluster theory calculations.
Our findings show that the improvements due to node-aware
topologies are less significant, but allow for more consistent performance across different node counts.
As the number of cores per node continues to grow on modern architectures, the benefit of node-aware mapping is likely to be more pronounced in the future.

\if
Future work will aim at further developments in CTF that
improve the performance of so-called
particle-particle ladder tensor contractions in coupled-cluster
calculations. Currently the performance in this contraction is not limited
by the communication volume in the tensor contraction operation but by
data redistribution, which could be handled in a more efficient manner.
\fi

\section{Acknowledgments}
Andreas Irmler and Andreas Grüneis acknowledge support from the European Union’s Horizon 2020 research and innovation program under Grant Agreement No. 951786 (The NOMAD CoE).
Raghavendra Kanakagiri and Edgar Solomonik received support from the US NSF OAC SSI program, via award No.\ 1931258.
The authors acknowledge application support and computing time of the MPCDF.





\bibliographystyle{splncs04}
\bibliography{nodeAware.bib}

\end{document}

%% file: result_memana.tex
\begingroup
  \makeatletter
  \providecommand\color[2][]{%
    \GenericError{(gnuplot) \space\space\space\@spaces}{%
      Package color not loaded in conjunction with
      terminal option `colourtext'%
    }{See the gnuplot documentation for explanation.%
    }{Either use 'blacktext' in gnuplot or load the package
      color.sty in LaTeX.}%
    \renewcommand\color[2][]{}%
  }%
  \providecommand\includegraphics[2][]{%
    \GenericError{(gnuplot) \space\space\space\@spaces}{%
      Package graphicx or graphics not loaded%
    }{See the gnuplot documentation for explanation.%
    }{The gnuplot epslatex terminal needs graphicx.sty or graphics.sty.}%
    \renewcommand\includegraphics[2][]{}%
  }%
  \providecommand\rotatebox[2]{#2}%
  \@ifundefined{ifGPcolor}{%
    \newif\ifGPcolor
    \GPcolorfalse
  }{}%
  \@ifundefined{ifGPblacktext}{%
    \newif\ifGPblacktext
    \GPblacktexttrue
  }{}%
  \let\gplgaddtomacro\g@addto@macro
  \gdef\gplbacktext{}%
  \gdef\gplfronttext{}%
  \makeatother
  \ifGPblacktext
    \def\colorrgb#1{}%
    \def\colorgray#1{}%
  \else
    \ifGPcolor
      \def\colorrgb#1{\color[rgb]{#1}}%
      \def\colorgray#1{\color[gray]{#1}}%
      \expandafter\def\csname LTw\endcsname{\color{white}}%
      \expandafter\def\csname LTb\endcsname{\color{black}}%
      \expandafter\def\csname LTa\endcsname{\color{black}}%
      \expandafter\def\csname LT0\endcsname{\color[rgb]{1,0,0}}%
      \expandafter\def\csname LT1\endcsname{\color[rgb]{0,1,0}}%
      \expandafter\def\csname LT2\endcsname{\color[rgb]{0,0,1}}%
      \expandafter\def\csname LT3\endcsname{\color[rgb]{1,0,1}}%
      \expandafter\def\csname LT4\endcsname{\color[rgb]{0,1,1}}%
      \expandafter\def\csname LT5\endcsname{\color[rgb]{1,1,0}}%
      \expandafter\def\csname LT6\endcsname{\color[rgb]{0,0,0}}%
      \expandafter\def\csname LT7\endcsname{\color[rgb]{1,0.3,0}}%
      \expandafter\def\csname LT8\endcsname{\color[rgb]{0.5,0.5,0.5}}%
    \else
      \def\colorrgb#1{\color{black}}%
      \def\colorgray#1{\color[gray]{#1}}%
      \expandafter\def\csname LTw\endcsname{\color{white}}%
      \expandafter\def\csname LTb\endcsname{\color{black}}%
      \expandafter\def\csname LTa\endcsname{\color{black}}%
      \expandafter\def\csname LT0\endcsname{\color{black}}%
      \expandafter\def\csname LT1\endcsname{\color{black}}%
      \expandafter\def\csname LT2\endcsname{\color{black}}%
      \expandafter\def\csname LT3\endcsname{\color{black}}%
      \expandafter\def\csname LT4\endcsname{\color{black}}%
      \expandafter\def\csname LT5\endcsname{\color{black}}%
      \expandafter\def\csname LT6\endcsname{\color{black}}%
      \expandafter\def\csname LT7\endcsname{\color{black}}%
      \expandafter\def\csname LT8\endcsname{\color{black}}%
    \fi
  \fi
    \setlength{\unitlength}{0.0500bp}%
    \ifx\gptboxheight\undefined%
      \newlength{\gptboxheight}%
      \newlength{\gptboxwidth}%
      \newsavebox{\gptboxtext}%
    \fi%
    \setlength{\fboxrule}{0.5pt}%
    \setlength{\fboxsep}{1pt}%
\begin{picture}(7200.00,2304.00)%
    \gplgaddtomacro\gplbacktext{%
      \csname LTb\endcsname
      \put(948,437){\makebox(0,0)[r]{\strut{}$0$}}%
      \csname LTb\endcsname
      \put(948,765){\makebox(0,0)[r]{\strut{}$5$}}%
      \csname LTb\endcsname
      \put(948,1094){\makebox(0,0)[r]{\strut{}$10$}}%
      \csname LTb\endcsname
      \put(948,1422){\makebox(0,0)[r]{\strut{}$15$}}%
      \csname LTb\endcsname
      \put(948,1750){\makebox(0,0)[r]{\strut{}$20$}}%
      \csname LTb\endcsname
      \put(1251,217){\makebox(0,0){\strut{}$1$}}%
      \csname LTb\endcsname
      \put(3363,217){\makebox(0,0){\strut{}$10$}}%
      \csname LTb\endcsname
      \put(5474,217){\makebox(0,0){\strut{}$100$}}%
    }%
    \gplgaddtomacro\gplfronttext{%
      \csname LTb\endcsname
      \put(475,1254){\rotatebox{-270}{\makebox(0,0){\strut{}HWM/storage}}}%
      \put(3851,-113){\makebox(0,0){\strut{}Number of nodes}}%
      \csname LTb\endcsname
      \put(1874,2179){\makebox(0,0)[r]{\strut{}square}}%
      \csname LTb\endcsname
      \put(3006,2179){\makebox(0,0)[r]{\strut{}largeK}}%
      \csname LTb\endcsname
      \put(4138,2179){\makebox(0,0)[r]{\strut{}largeM}}%
      \csname LTb\endcsname
      \put(5270,2179){\makebox(0,0)[r]{\strut{}smallK}}%
    }%
    \gplgaddtomacro\gplbacktext{%
      \csname LTb\endcsname
      \put(948,437){\makebox(0,0)[r]{\strut{}$0$}}%
      \csname LTb\endcsname
      \put(948,765){\makebox(0,0)[r]{\strut{}$5$}}%
      \csname LTb\endcsname
      \put(948,1094){\makebox(0,0)[r]{\strut{}$10$}}%
      \csname LTb\endcsname
      \put(948,1422){\makebox(0,0)[r]{\strut{}$15$}}%
      \csname LTb\endcsname
      \put(948,1750){\makebox(0,0)[r]{\strut{}$20$}}%
      \csname LTb\endcsname
      \put(1251,217){\makebox(0,0){\strut{}$1$}}%
      \csname LTb\endcsname
      \put(3363,217){\makebox(0,0){\strut{}$10$}}%
      \csname LTb\endcsname
      \put(5474,217){\makebox(0,0){\strut{}$100$}}%
    }%
    \gplgaddtomacro\gplfronttext{%
      \csname LTb\endcsname
      \put(475,1254){\rotatebox{-270}{\makebox(0,0){\strut{}HWM/storage}}}%
      \put(3851,-113){\makebox(0,0){\strut{}Number of nodes}}%
      \csname LTb\endcsname
      \put(4755,1962){\makebox(0,0)[r]{\strut{}CTF}}%
      \csname LTb\endcsname
      \put(4755,1742){\makebox(0,0)[r]{\strut{}ScaLAPACK}}%
      \csname LTb\endcsname
      \put(6283,1962){\makebox(0,0)[r]{\strut{}COSMA-unl}}%
      \csname LTb\endcsname
      \put(6283,1742){\makebox(0,0)[r]{\strut{}COSMA-lim}}%
    }%
    \gplbacktext
    \put(0,0){\includegraphics{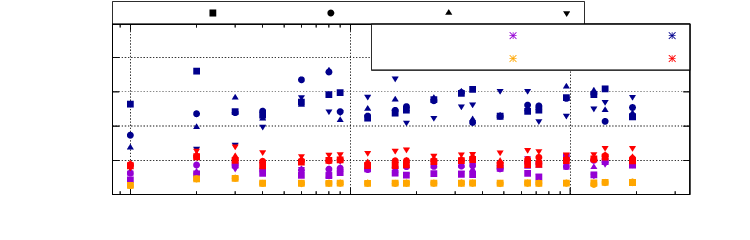}}%
    \gplfronttext
  \end{picture}%
\endgroup

%% file: result_matrix.tex
\begingroup
  \makeatletter
  \providecommand\color[2][]{%
    \GenericError{(gnuplot) \space\space\space\@spaces}{%
      Package color not loaded in conjunction with
      terminal option `colourtext'%
    }{See the gnuplot documentation for explanation.%
    }{Either use 'blacktext' in gnuplot or load the package
      color.sty in LaTeX.}%
    \renewcommand\color[2][]{}%
  }%
  \providecommand\includegraphics[2][]{%
    \GenericError{(gnuplot) \space\space\space\@spaces}{%
      Package graphicx or graphics not loaded%
    }{See the gnuplot documentation for explanation.%
    }{The gnuplot epslatex terminal needs graphicx.sty or graphics.sty.}%
    \renewcommand\includegraphics[2][]{}%
  }%
  \providecommand\rotatebox[2]{#2}%
  \@ifundefined{ifGPcolor}{%
    \newif\ifGPcolor
    \GPcolorfalse
  }{}%
  \@ifundefined{ifGPblacktext}{%
    \newif\ifGPblacktext
    \GPblacktexttrue
  }{}%
  \let\gplgaddtomacro\g@addto@macro
  \gdef\gplbacktext{}%
  \gdef\gplfronttext{}%
  \makeatother
  \ifGPblacktext
    \def\colorrgb#1{}%
    \def\colorgray#1{}%
  \else
    \ifGPcolor
      \def\colorrgb#1{\color[rgb]{#1}}%
      \def\colorgray#1{\color[gray]{#1}}%
      \expandafter\def\csname LTw\endcsname{\color{white}}%
      \expandafter\def\csname LTb\endcsname{\color{black}}%
      \expandafter\def\csname LTa\endcsname{\color{black}}%
      \expandafter\def\csname LT0\endcsname{\color[rgb]{1,0,0}}%
      \expandafter\def\csname LT1\endcsname{\color[rgb]{0,1,0}}%
      \expandafter\def\csname LT2\endcsname{\color[rgb]{0,0,1}}%
      \expandafter\def\csname LT3\endcsname{\color[rgb]{1,0,1}}%
      \expandafter\def\csname LT4\endcsname{\color[rgb]{0,1,1}}%
      \expandafter\def\csname LT5\endcsname{\color[rgb]{1,1,0}}%
      \expandafter\def\csname LT6\endcsname{\color[rgb]{0,0,0}}%
      \expandafter\def\csname LT7\endcsname{\color[rgb]{1,0.3,0}}%
      \expandafter\def\csname LT8\endcsname{\color[rgb]{0.5,0.5,0.5}}%
    \else
      \def\colorrgb#1{\color{black}}%
      \def\colorgray#1{\color[gray]{#1}}%
      \expandafter\def\csname LTw\endcsname{\color{white}}%
      \expandafter\def\csname LTb\endcsname{\color{black}}%
      \expandafter\def\csname LTa\endcsname{\color{black}}%
      \expandafter\def\csname LT0\endcsname{\color{black}}%
      \expandafter\def\csname LT1\endcsname{\color{black}}%
      \expandafter\def\csname LT2\endcsname{\color{black}}%
      \expandafter\def\csname LT3\endcsname{\color{black}}%
      \expandafter\def\csname LT4\endcsname{\color{black}}%
      \expandafter\def\csname LT5\endcsname{\color{black}}%
      \expandafter\def\csname LT6\endcsname{\color{black}}%
      \expandafter\def\csname LT7\endcsname{\color{black}}%
      \expandafter\def\csname LT8\endcsname{\color{black}}%
    \fi
  \fi
    \setlength{\unitlength}{0.0500bp}%
    \ifx\gptboxheight\undefined%
      \newlength{\gptboxheight}%
      \newlength{\gptboxwidth}%
      \newsavebox{\gptboxtext}%
    \fi%
    \setlength{\fboxrule}{0.5pt}%
    \setlength{\fboxsep}{1pt}%
\begin{picture}(5760.00,9360.00)%
    \gplgaddtomacro\gplbacktext{%
      \csname LTb\endcsname
      \put(-75,7406){\makebox(0,0)[r]{\strut{}$0$}}%
      \csname LTb\endcsname
      \put(-75,7676){\makebox(0,0)[r]{\strut{}$10$}}%
      \csname LTb\endcsname
      \put(-75,7946){\makebox(0,0)[r]{\strut{}$20$}}%
      \csname LTb\endcsname
      \put(-75,8216){\makebox(0,0)[r]{\strut{}$30$}}%
      \csname LTb\endcsname
      \put(-75,8486){\makebox(0,0)[r]{\strut{}$40$}}%
      \csname LTb\endcsname
      \put(-75,8756){\makebox(0,0)[r]{\strut{}$50$}}%
      \csname LTb\endcsname
      \put(106,7186){\makebox(0,0){\strut{}$1$}}%
      \csname LTb\endcsname
      \put(705,7186){\makebox(0,0){\strut{}$10$}}%
      \csname LTb\endcsname
      \put(1305,7186){\makebox(0,0){\strut{}$100$}}%
    }%
    \gplgaddtomacro\gplfronttext{%
      \csname LTb\endcsname
      \put(211,7568){\makebox(0,0){\strut{}A}}%
      \put(497,7838){\makebox(0,0){\strut{}B}}%
      \put(497,7568){\makebox(0,0){\strut{}C}}%
    }%
    \gplgaddtomacro\gplbacktext{%
      \csname LTb\endcsname
      \put(1960,7406){\makebox(0,0)[r]{\strut{}$0$}}%
      \csname LTb\endcsname
      \put(1960,7676){\makebox(0,0)[r]{\strut{}$10$}}%
      \csname LTb\endcsname
      \put(1960,7946){\makebox(0,0)[r]{\strut{}$20$}}%
      \csname LTb\endcsname
      \put(1960,8216){\makebox(0,0)[r]{\strut{}$30$}}%
      \csname LTb\endcsname
      \put(1960,8486){\makebox(0,0)[r]{\strut{}$40$}}%
      \csname LTb\endcsname
      \put(1960,8756){\makebox(0,0)[r]{\strut{}$50$}}%
      \csname LTb\endcsname
      \put(2141,7186){\makebox(0,0){\strut{}$1$}}%
      \csname LTb\endcsname
      \put(2740,7186){\makebox(0,0){\strut{}$10$}}%
      \csname LTb\endcsname
      \put(3340,7186){\makebox(0,0){\strut{}$100$}}%
    }%
    \gplgaddtomacro\gplfronttext{%
    }%
    \gplgaddtomacro\gplbacktext{%
      \csname LTb\endcsname
      \put(3996,7406){\makebox(0,0)[r]{\strut{}$0$}}%
      \csname LTb\endcsname
      \put(3996,7676){\makebox(0,0)[r]{\strut{}$10$}}%
      \csname LTb\endcsname
      \put(3996,7946){\makebox(0,0)[r]{\strut{}$20$}}%
      \csname LTb\endcsname
      \put(3996,8216){\makebox(0,0)[r]{\strut{}$30$}}%
      \csname LTb\endcsname
      \put(3996,8486){\makebox(0,0)[r]{\strut{}$40$}}%
      \csname LTb\endcsname
      \put(3996,8756){\makebox(0,0)[r]{\strut{}$50$}}%
      \csname LTb\endcsname
      \put(4176,7186){\makebox(0,0){\strut{}$1$}}%
      \csname LTb\endcsname
      \put(4776,7186){\makebox(0,0){\strut{}$10$}}%
      \csname LTb\endcsname
      \put(5375,7186){\makebox(0,0){\strut{}$100$}}%
    }%
    \gplgaddtomacro\gplfronttext{%
    }%
    \gplgaddtomacro\gplbacktext{%
      \csname LTb\endcsname
      \put(-75,5498){\makebox(0,0)[r]{\strut{}$0$}}%
      \csname LTb\endcsname
      \put(-75,5768){\makebox(0,0)[r]{\strut{}$10$}}%
      \csname LTb\endcsname
      \put(-75,6038){\makebox(0,0)[r]{\strut{}$20$}}%
      \csname LTb\endcsname
      \put(-75,6309){\makebox(0,0)[r]{\strut{}$30$}}%
      \csname LTb\endcsname
      \put(-75,6579){\makebox(0,0)[r]{\strut{}$40$}}%
      \csname LTb\endcsname
      \put(-75,6849){\makebox(0,0)[r]{\strut{}$50$}}%
      \csname LTb\endcsname
      \put(106,5278){\makebox(0,0){\strut{}$1$}}%
      \csname LTb\endcsname
      \put(705,5278){\makebox(0,0){\strut{}$10$}}%
      \csname LTb\endcsname
      \put(1305,5278){\makebox(0,0){\strut{}$100$}}%
    }%
    \gplgaddtomacro\gplfronttext{%
      \csname LTb\endcsname
      \put(273,5622){\makebox(0,0){\strut{}A}}%
      \put(545,5876){\makebox(0,0){\strut{}B}}%
      \put(545,5622){\makebox(0,0){\strut{}C}}%
    }%
    \gplgaddtomacro\gplbacktext{%
      \csname LTb\endcsname
      \put(1960,5498){\makebox(0,0)[r]{\strut{}$0$}}%
      \csname LTb\endcsname
      \put(1960,5768){\makebox(0,0)[r]{\strut{}$10$}}%
      \csname LTb\endcsname
      \put(1960,6038){\makebox(0,0)[r]{\strut{}$20$}}%
      \csname LTb\endcsname
      \put(1960,6309){\makebox(0,0)[r]{\strut{}$30$}}%
      \csname LTb\endcsname
      \put(1960,6579){\makebox(0,0)[r]{\strut{}$40$}}%
      \csname LTb\endcsname
      \put(1960,6849){\makebox(0,0)[r]{\strut{}$50$}}%
      \csname LTb\endcsname
      \put(2141,5278){\makebox(0,0){\strut{}$1$}}%
      \csname LTb\endcsname
      \put(2740,5278){\makebox(0,0){\strut{}$10$}}%
      \csname LTb\endcsname
      \put(3340,5278){\makebox(0,0){\strut{}$100$}}%
    }%
    \gplgaddtomacro\gplfronttext{%
    }%
    \gplgaddtomacro\gplbacktext{%
      \csname LTb\endcsname
      \put(3996,5498){\makebox(0,0)[r]{\strut{}$0$}}%
      \csname LTb\endcsname
      \put(3996,5768){\makebox(0,0)[r]{\strut{}$10$}}%
      \csname LTb\endcsname
      \put(3996,6038){\makebox(0,0)[r]{\strut{}$20$}}%
      \csname LTb\endcsname
      \put(3996,6309){\makebox(0,0)[r]{\strut{}$30$}}%
      \csname LTb\endcsname
      \put(3996,6579){\makebox(0,0)[r]{\strut{}$40$}}%
      \csname LTb\endcsname
      \put(3996,6849){\makebox(0,0)[r]{\strut{}$50$}}%
      \csname LTb\endcsname
      \put(4176,5278){\makebox(0,0){\strut{}$1$}}%
      \csname LTb\endcsname
      \put(4776,5278){\makebox(0,0){\strut{}$10$}}%
      \csname LTb\endcsname
      \put(5375,5278){\makebox(0,0){\strut{}$100$}}%
    }%
    \gplgaddtomacro\gplfronttext{%
    }%
    \gplgaddtomacro\gplbacktext{%
      \csname LTb\endcsname
      \put(-75,3591){\makebox(0,0)[r]{\strut{}$0$}}%
      \csname LTb\endcsname
      \put(-75,3861){\makebox(0,0)[r]{\strut{}$10$}}%
      \csname LTb\endcsname
      \put(-75,4131){\makebox(0,0)[r]{\strut{}$20$}}%
      \csname LTb\endcsname
      \put(-75,4402){\makebox(0,0)[r]{\strut{}$30$}}%
      \csname LTb\endcsname
      \put(-75,4672){\makebox(0,0)[r]{\strut{}$40$}}%
      \csname LTb\endcsname
      \put(-75,4942){\makebox(0,0)[r]{\strut{}$50$}}%
      \csname LTb\endcsname
      \put(106,3371){\makebox(0,0){\strut{}$1$}}%
      \csname LTb\endcsname
      \put(705,3371){\makebox(0,0){\strut{}$10$}}%
      \csname LTb\endcsname
      \put(1305,3371){\makebox(0,0){\strut{}$100$}}%
    }%
    \gplgaddtomacro\gplfronttext{%
      \csname LTb\endcsname
      \put(174,3794){\makebox(0,0){\strut{}A}}%
      \put(374,4077){\makebox(0,0){\strut{}B}}%
      \put(374,3794){\makebox(0,0){\strut{}C}}%
    }%
    \gplgaddtomacro\gplbacktext{%
      \csname LTb\endcsname
      \put(1960,3591){\makebox(0,0)[r]{\strut{}$0$}}%
      \csname LTb\endcsname
      \put(1960,3861){\makebox(0,0)[r]{\strut{}$10$}}%
      \csname LTb\endcsname
      \put(1960,4131){\makebox(0,0)[r]{\strut{}$20$}}%
      \csname LTb\endcsname
      \put(1960,4402){\makebox(0,0)[r]{\strut{}$30$}}%
      \csname LTb\endcsname
      \put(1960,4672){\makebox(0,0)[r]{\strut{}$40$}}%
      \csname LTb\endcsname
      \put(1960,4942){\makebox(0,0)[r]{\strut{}$50$}}%
      \csname LTb\endcsname
      \put(2141,3371){\makebox(0,0){\strut{}$1$}}%
      \csname LTb\endcsname
      \put(2740,3371){\makebox(0,0){\strut{}$10$}}%
      \csname LTb\endcsname
      \put(3340,3371){\makebox(0,0){\strut{}$100$}}%
    }%
    \gplgaddtomacro\gplfronttext{%
    }%
    \gplgaddtomacro\gplbacktext{%
      \csname LTb\endcsname
      \put(3996,3591){\makebox(0,0)[r]{\strut{}$0$}}%
      \csname LTb\endcsname
      \put(3996,3861){\makebox(0,0)[r]{\strut{}$10$}}%
      \csname LTb\endcsname
      \put(3996,4131){\makebox(0,0)[r]{\strut{}$20$}}%
      \csname LTb\endcsname
      \put(3996,4402){\makebox(0,0)[r]{\strut{}$30$}}%
      \csname LTb\endcsname
      \put(3996,4672){\makebox(0,0)[r]{\strut{}$40$}}%
      \csname LTb\endcsname
      \put(3996,4942){\makebox(0,0)[r]{\strut{}$50$}}%
      \csname LTb\endcsname
      \put(4176,3371){\makebox(0,0){\strut{}$1$}}%
      \csname LTb\endcsname
      \put(4776,3371){\makebox(0,0){\strut{}$10$}}%
      \csname LTb\endcsname
      \put(5375,3371){\makebox(0,0){\strut{}$100$}}%
    }%
    \gplgaddtomacro\gplfronttext{%
    }%
    \gplgaddtomacro\gplbacktext{%
      \csname LTb\endcsname
      \put(-75,1684){\makebox(0,0)[r]{\strut{}$0$}}%
      \csname LTb\endcsname
      \put(-75,1954){\makebox(0,0)[r]{\strut{}$10$}}%
      \csname LTb\endcsname
      \put(-75,2224){\makebox(0,0)[r]{\strut{}$20$}}%
      \csname LTb\endcsname
      \put(-75,2495){\makebox(0,0)[r]{\strut{}$30$}}%
      \csname LTb\endcsname
      \put(-75,2765){\makebox(0,0)[r]{\strut{}$40$}}%
      \csname LTb\endcsname
      \put(-75,3035){\makebox(0,0)[r]{\strut{}$50$}}%
      \csname LTb\endcsname
      \put(106,1464){\makebox(0,0){\strut{}$1$}}%
      \csname LTb\endcsname
      \put(705,1464){\makebox(0,0){\strut{}$10$}}%
      \csname LTb\endcsname
      \put(1305,1464){\makebox(0,0){\strut{}$100$}}%
    }%
    \gplgaddtomacro\gplfronttext{%
      \csname LTb\endcsname
      \put(174,1887){\makebox(0,0){\strut{}A}}%
      \put(473,2170){\makebox(0,0){\strut{}B}}%
      \put(473,1887){\makebox(0,0){\strut{}C}}%
    }%
    \gplgaddtomacro\gplbacktext{%
      \csname LTb\endcsname
      \put(1960,1684){\makebox(0,0)[r]{\strut{}$0$}}%
      \csname LTb\endcsname
      \put(1960,1954){\makebox(0,0)[r]{\strut{}$10$}}%
      \csname LTb\endcsname
      \put(1960,2224){\makebox(0,0)[r]{\strut{}$20$}}%
      \csname LTb\endcsname
      \put(1960,2495){\makebox(0,0)[r]{\strut{}$30$}}%
      \csname LTb\endcsname
      \put(1960,2765){\makebox(0,0)[r]{\strut{}$40$}}%
      \csname LTb\endcsname
      \put(1960,3035){\makebox(0,0)[r]{\strut{}$50$}}%
      \csname LTb\endcsname
      \put(2141,1464){\makebox(0,0){\strut{}$1$}}%
      \csname LTb\endcsname
      \put(2740,1464){\makebox(0,0){\strut{}$10$}}%
      \csname LTb\endcsname
      \put(3340,1464){\makebox(0,0){\strut{}$100$}}%
    }%
    \gplgaddtomacro\gplfronttext{%
    }%
    \gplgaddtomacro\gplbacktext{%
      \csname LTb\endcsname
      \put(3996,1684){\makebox(0,0)[r]{\strut{}$0$}}%
      \csname LTb\endcsname
      \put(3996,1954){\makebox(0,0)[r]{\strut{}$10$}}%
      \csname LTb\endcsname
      \put(3996,2224){\makebox(0,0)[r]{\strut{}$20$}}%
      \csname LTb\endcsname
      \put(3996,2495){\makebox(0,0)[r]{\strut{}$30$}}%
      \csname LTb\endcsname
      \put(3996,2765){\makebox(0,0)[r]{\strut{}$40$}}%
      \csname LTb\endcsname
      \put(3996,3035){\makebox(0,0)[r]{\strut{}$50$}}%
      \csname LTb\endcsname
      \put(4176,1464){\makebox(0,0){\strut{}$1$}}%
      \csname LTb\endcsname
      \put(4776,1464){\makebox(0,0){\strut{}$10$}}%
      \csname LTb\endcsname
      \put(5375,1464){\makebox(0,0){\strut{}$100$}}%
      \put(581,9033){\makebox(0,0)[l]{\strut{}strong}}%
      \put(2379,9033){\makebox(0,0)[l]{\strut{}weak 18MB}}%
      \put(4382,9033){\makebox(0,0)[l]{\strut{}weak 180MB}}%
      \put(-448,1900){\rotatebox{90}{\makebox(0,0)[l]{\strut{}GFLOPS/core}}}%
      \put(-448,3845){\rotatebox{90}{\makebox(0,0)[l]{\strut{}GFLOPS/core}}}%
      \put(-448,5791){\rotatebox{90}{\makebox(0,0)[l]{\strut{}GFLOPS/core}}}%
      \put(-448,7763){\rotatebox{90}{\makebox(0,0)[l]{\strut{}GFLOPS/core}}}%
      \put(4176,1279){\makebox(0,0)[l]{\strut{}Number of Nodes}}%
      \put(2065,1279){\makebox(0,0)[l]{\strut{}Number of Nodes}}%
      \put(-16,1279){\makebox(0,0)[l]{\strut{}Number of Nodes}}%
    }%
    \gplgaddtomacro\gplfronttext{%
      \csname LTb\endcsname
      \put(1774,899){\makebox(0,0)[r]{\strut{}CTF-def}}%
      \csname LTb\endcsname
      \put(1774,679){\makebox(0,0)[r]{\strut{}CTF-na}}%
      \csname LTb\endcsname
      \put(3291,899){\makebox(0,0)[r]{\strut{}COSMA-unl}}%
      \csname LTb\endcsname
      \put(3291,679){\makebox(0,0)[r]{\strut{}COSMA-lim}}%
      \csname LTb\endcsname
      \put(4808,899){\makebox(0,0)[r]{\strut{}ScaLAPACK}}%
    }%
    \gplbacktext
    \put(0,0){\includegraphics{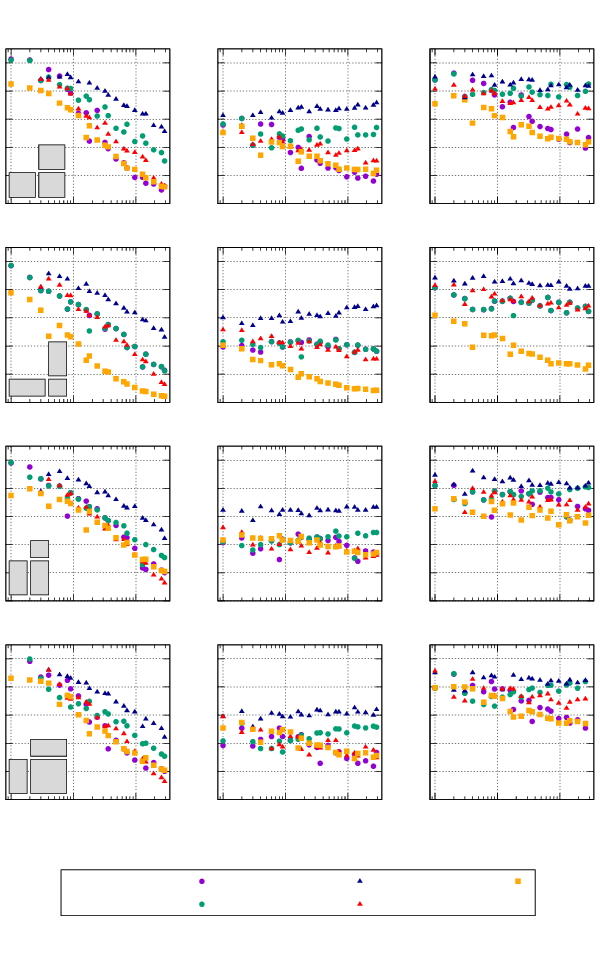}}%
    \gplfronttext
  \end{picture}%
\endgroup

%% file: result_stampede.tex
\begingroup
  \makeatletter
  \providecommand\color[2][]{%
    \GenericError{(gnuplot) \space\space\space\@spaces}{%
      Package color not loaded in conjunction with
      terminal option `colourtext'%
    }{See the gnuplot documentation for explanation.%
    }{Either use 'blacktext' in gnuplot or load the package
      color.sty in LaTeX.}%
    \renewcommand\color[2][]{}%
  }%
  \providecommand\includegraphics[2][]{%
    \GenericError{(gnuplot) \space\space\space\@spaces}{%
      Package graphicx or graphics not loaded%
    }{See the gnuplot documentation for explanation.%
    }{The gnuplot epslatex terminal needs graphicx.sty or graphics.sty.}%
    \renewcommand\includegraphics[2][]{}%
  }%
  \providecommand\rotatebox[2]{#2}%
  \@ifundefined{ifGPcolor}{%
    \newif\ifGPcolor
    \GPcolorfalse
  }{}%
  \@ifundefined{ifGPblacktext}{%
    \newif\ifGPblacktext
    \GPblacktexttrue
  }{}%
  \let\gplgaddtomacro\g@addto@macro
  \gdef\gplbacktext{}%
  \gdef\gplfronttext{}%
  \makeatother
  \ifGPblacktext
    \def\colorrgb#1{}%
    \def\colorgray#1{}%
  \else
    \ifGPcolor
      \def\colorrgb#1{\color[rgb]{#1}}%
      \def\colorgray#1{\color[gray]{#1}}%
      \expandafter\def\csname LTw\endcsname{\color{white}}%
      \expandafter\def\csname LTb\endcsname{\color{black}}%
      \expandafter\def\csname LTa\endcsname{\color{black}}%
      \expandafter\def\csname LT0\endcsname{\color[rgb]{1,0,0}}%
      \expandafter\def\csname LT1\endcsname{\color[rgb]{0,1,0}}%
      \expandafter\def\csname LT2\endcsname{\color[rgb]{0,0,1}}%
      \expandafter\def\csname LT3\endcsname{\color[rgb]{1,0,1}}%
      \expandafter\def\csname LT4\endcsname{\color[rgb]{0,1,1}}%
      \expandafter\def\csname LT5\endcsname{\color[rgb]{1,1,0}}%
      \expandafter\def\csname LT6\endcsname{\color[rgb]{0,0,0}}%
      \expandafter\def\csname LT7\endcsname{\color[rgb]{1,0.3,0}}%
      \expandafter\def\csname LT8\endcsname{\color[rgb]{0.5,0.5,0.5}}%
    \else
      \def\colorrgb#1{\color{black}}%
      \def\colorgray#1{\color[gray]{#1}}%
      \expandafter\def\csname LTw\endcsname{\color{white}}%
      \expandafter\def\csname LTb\endcsname{\color{black}}%
      \expandafter\def\csname LTa\endcsname{\color{black}}%
      \expandafter\def\csname LT0\endcsname{\color{black}}%
      \expandafter\def\csname LT1\endcsname{\color{black}}%
      \expandafter\def\csname LT2\endcsname{\color{black}}%
      \expandafter\def\csname LT3\endcsname{\color{black}}%
      \expandafter\def\csname LT4\endcsname{\color{black}}%
      \expandafter\def\csname LT5\endcsname{\color{black}}%
      \expandafter\def\csname LT6\endcsname{\color{black}}%
      \expandafter\def\csname LT7\endcsname{\color{black}}%
      \expandafter\def\csname LT8\endcsname{\color{black}}%
    \fi
  \fi
    \setlength{\unitlength}{0.0500bp}%
    \ifx\gptboxheight\undefined%
      \newlength{\gptboxheight}%
      \newlength{\gptboxwidth}%
      \newsavebox{\gptboxtext}%
    \fi%
    \setlength{\fboxrule}{0.5pt}%
    \setlength{\fboxsep}{1pt}%
\begin{picture}(5760.00,2160.00)%
    \gplgaddtomacro\gplbacktext{%
      \csname LTb\endcsname
      \put(-132,432){\makebox(0,0)[r]{\strut{}$0$}}%
      \csname LTb\endcsname
      \put(-132,914){\makebox(0,0)[r]{\strut{}$10$}}%
      \csname LTb\endcsname
      \put(-132,1395){\makebox(0,0)[r]{\strut{}$20$}}%
      \csname LTb\endcsname
      \put(36,212){\makebox(0,0){\strut{}$1$}}%
      \csname LTb\endcsname
      \put(486,212){\makebox(0,0){\strut{}$10$}}%
      \csname LTb\endcsname
      \put(935,212){\makebox(0,0){\strut{}$100$}}%
    }%
    \gplgaddtomacro\gplfronttext{%
    }%
    \gplgaddtomacro\gplbacktext{%
      \csname LTb\endcsname
      \put(1394,432){\makebox(0,0)[r]{\strut{}$0$}}%
      \csname LTb\endcsname
      \put(1394,914){\makebox(0,0)[r]{\strut{}$10$}}%
      \csname LTb\endcsname
      \put(1394,1395){\makebox(0,0)[r]{\strut{}$20$}}%
      \csname LTb\endcsname
      \put(1562,212){\makebox(0,0){\strut{}$1$}}%
      \csname LTb\endcsname
      \put(2012,212){\makebox(0,0){\strut{}$10$}}%
      \csname LTb\endcsname
      \put(2461,212){\makebox(0,0){\strut{}$100$}}%
    }%
    \gplgaddtomacro\gplfronttext{%
    }%
    \gplgaddtomacro\gplbacktext{%
      \csname LTb\endcsname
      \put(2920,432){\makebox(0,0)[r]{\strut{}$0$}}%
      \csname LTb\endcsname
      \put(2920,914){\makebox(0,0)[r]{\strut{}$10$}}%
      \csname LTb\endcsname
      \put(2920,1395){\makebox(0,0)[r]{\strut{}$20$}}%
      \csname LTb\endcsname
      \put(3088,212){\makebox(0,0){\strut{}$1$}}%
      \csname LTb\endcsname
      \put(3538,212){\makebox(0,0){\strut{}$10$}}%
      \csname LTb\endcsname
      \put(3987,212){\makebox(0,0){\strut{}$100$}}%
    }%
    \gplgaddtomacro\gplfronttext{%
    }%
    \gplgaddtomacro\gplbacktext{%
      \csname LTb\endcsname
      \put(4447,432){\makebox(0,0)[r]{\strut{}$0$}}%
      \csname LTb\endcsname
      \put(4447,914){\makebox(0,0)[r]{\strut{}$10$}}%
      \csname LTb\endcsname
      \put(4447,1395){\makebox(0,0)[r]{\strut{}$20$}}%
      \csname LTb\endcsname
      \put(4615,212){\makebox(0,0){\strut{}$1$}}%
      \csname LTb\endcsname
      \put(5065,212){\makebox(0,0){\strut{}$10$}}%
      \csname LTb\endcsname
      \put(5514,212){\makebox(0,0){\strut{}$100$}}%
      \put(-463,480){\rotatebox{90}{\makebox(0,0)[l]{\strut{}GFLOPS/core}}}%
      \put(4615,-48){\makebox(0,0)[l]{\strut{}Number of Nodes}}%
      \put(3032,-48){\makebox(0,0)[l]{\strut{}Number of Nodes}}%
      \put(1469,-48){\makebox(0,0)[l]{\strut{}Number of Nodes}}%
      \put(-113,-48){\makebox(0,0)[l]{\strut{}Number of Nodes}}%
    }%
    \gplgaddtomacro\gplfronttext{%
      \csname LTb\endcsname
      \put(1973,2031){\makebox(0,0)[r]{\strut{}CTF-def}}%
      \csname LTb\endcsname
      \put(1973,1811){\makebox(0,0)[r]{\strut{}CTF-na}}%
      \csname LTb\endcsname
      \put(3490,2031){\makebox(0,0)[r]{\strut{}COSMA-unl}}%
      \csname LTb\endcsname
      \put(3490,1811){\makebox(0,0)[r]{\strut{}COSMA-lim}}%
      \csname LTb\endcsname
      \put(5007,2031){\makebox(0,0)[r]{\strut{}ScaLAPACK}}%
    }%
    \gplbacktext
    \put(0,0){\includegraphics{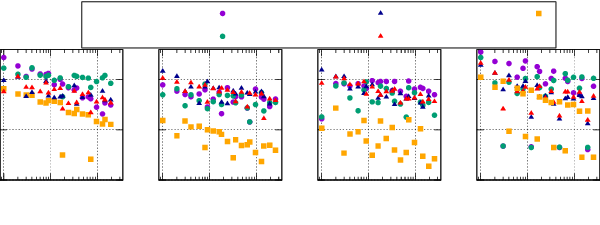}}%
    \gplfronttext
  \end{picture}%
\endgroup

%% file: result_ring.tex
\begingroup
  \makeatletter
  \providecommand\color[2][]{%
    \GenericError{(gnuplot) \space\space\space\@spaces}{%
      Package color not loaded in conjunction with
      terminal option `colourtext'%
    }{See the gnuplot documentation for explanation.%
    }{Either use 'blacktext' in gnuplot or load the package
      color.sty in LaTeX.}%
    \renewcommand\color[2][]{}%
  }%
  \providecommand\includegraphics[2][]{%
    \GenericError{(gnuplot) \space\space\space\@spaces}{%
      Package graphicx or graphics not loaded%
    }{See the gnuplot documentation for explanation.%
    }{The gnuplot epslatex terminal needs graphicx.sty or graphics.sty.}%
    \renewcommand\includegraphics[2][]{}%
  }%
  \providecommand\rotatebox[2]{#2}%
  \@ifundefined{ifGPcolor}{%
    \newif\ifGPcolor
    \GPcolorfalse
  }{}%
  \@ifundefined{ifGPblacktext}{%
    \newif\ifGPblacktext
    \GPblacktexttrue
  }{}%
  \let\gplgaddtomacro\g@addto@macro
  \gdef\gplbacktext{}%
  \gdef\gplfronttext{}%
  \makeatother
  \ifGPblacktext
    \def\colorrgb#1{}%
    \def\colorgray#1{}%
  \else
    \ifGPcolor
      \def\colorrgb#1{\color[rgb]{#1}}%
      \def\colorgray#1{\color[gray]{#1}}%
      \expandafter\def\csname LTw\endcsname{\color{white}}%
      \expandafter\def\csname LTb\endcsname{\color{black}}%
      \expandafter\def\csname LTa\endcsname{\color{black}}%
      \expandafter\def\csname LT0\endcsname{\color[rgb]{1,0,0}}%
      \expandafter\def\csname LT1\endcsname{\color[rgb]{0,1,0}}%
      \expandafter\def\csname LT2\endcsname{\color[rgb]{0,0,1}}%
      \expandafter\def\csname LT3\endcsname{\color[rgb]{1,0,1}}%
      \expandafter\def\csname LT4\endcsname{\color[rgb]{0,1,1}}%
      \expandafter\def\csname LT5\endcsname{\color[rgb]{1,1,0}}%
      \expandafter\def\csname LT6\endcsname{\color[rgb]{0,0,0}}%
      \expandafter\def\csname LT7\endcsname{\color[rgb]{1,0.3,0}}%
      \expandafter\def\csname LT8\endcsname{\color[rgb]{0.5,0.5,0.5}}%
    \else
      \def\colorrgb#1{\color{black}}%
      \def\colorgray#1{\color[gray]{#1}}%
      \expandafter\def\csname LTw\endcsname{\color{white}}%
      \expandafter\def\csname LTb\endcsname{\color{black}}%
      \expandafter\def\csname LTa\endcsname{\color{black}}%
      \expandafter\def\csname LT0\endcsname{\color{black}}%
      \expandafter\def\csname LT1\endcsname{\color{black}}%
      \expandafter\def\csname LT2\endcsname{\color{black}}%
      \expandafter\def\csname LT3\endcsname{\color{black}}%
      \expandafter\def\csname LT4\endcsname{\color{black}}%
      \expandafter\def\csname LT5\endcsname{\color{black}}%
      \expandafter\def\csname LT6\endcsname{\color{black}}%
      \expandafter\def\csname LT7\endcsname{\color{black}}%
      \expandafter\def\csname LT8\endcsname{\color{black}}%
    \fi
  \fi
    \setlength{\unitlength}{0.0500bp}%
    \ifx\gptboxheight\undefined%
      \newlength{\gptboxheight}%
      \newlength{\gptboxwidth}%
      \newsavebox{\gptboxtext}%
    \fi%
    \setlength{\fboxrule}{0.5pt}%
    \setlength{\fboxsep}{1pt}%
\begin{picture}(5760.00,1872.00)%
    \gplgaddtomacro\gplbacktext{%
      \csname LTb\endcsname
      \put(-75,374){\makebox(0,0)[r]{\strut{}$0$}}%
      \csname LTb\endcsname
      \put(-75,612){\makebox(0,0)[r]{\strut{}$10$}}%
      \csname LTb\endcsname
      \put(-75,850){\makebox(0,0)[r]{\strut{}$20$}}%
      \csname LTb\endcsname
      \put(-75,1088){\makebox(0,0)[r]{\strut{}$30$}}%
      \csname LTb\endcsname
      \put(-75,1326){\makebox(0,0)[r]{\strut{}$40$}}%
      \csname LTb\endcsname
      \put(-75,1564){\makebox(0,0)[r]{\strut{}$50$}}%
      \csname LTb\endcsname
      \put(106,154){\makebox(0,0){\strut{}$1$}}%
      \csname LTb\endcsname
      \put(705,154){\makebox(0,0){\strut{}$10$}}%
      \csname LTb\endcsname
      \put(1305,154){\makebox(0,0){\strut{}$100$}}%
    }%
    \gplgaddtomacro\gplfronttext{%
      \csname LTb\endcsname
      \put(844,-176){\makebox(0,0){\strut{}Number of Nodes}}%
    }%
    \gplgaddtomacro\gplbacktext{%
      \csname LTb\endcsname
      \put(1960,374){\makebox(0,0)[r]{\strut{}$0$}}%
      \csname LTb\endcsname
      \put(1960,612){\makebox(0,0)[r]{\strut{}$10$}}%
      \csname LTb\endcsname
      \put(1960,850){\makebox(0,0)[r]{\strut{}$20$}}%
      \csname LTb\endcsname
      \put(1960,1088){\makebox(0,0)[r]{\strut{}$30$}}%
      \csname LTb\endcsname
      \put(1960,1326){\makebox(0,0)[r]{\strut{}$40$}}%
      \csname LTb\endcsname
      \put(1960,1564){\makebox(0,0)[r]{\strut{}$50$}}%
      \csname LTb\endcsname
      \put(2141,154){\makebox(0,0){\strut{}$1$}}%
      \csname LTb\endcsname
      \put(2740,154){\makebox(0,0){\strut{}$10$}}%
      \csname LTb\endcsname
      \put(3340,154){\makebox(0,0){\strut{}$100$}}%
    }%
    \gplgaddtomacro\gplfronttext{%
      \csname LTb\endcsname
      \put(2879,-176){\makebox(0,0){\strut{}Number of Nodes}}%
    }%
    \gplgaddtomacro\gplbacktext{%
      \csname LTb\endcsname
      \put(3996,374){\makebox(0,0)[r]{\strut{}$0$}}%
      \csname LTb\endcsname
      \put(3996,612){\makebox(0,0)[r]{\strut{}$10$}}%
      \csname LTb\endcsname
      \put(3996,850){\makebox(0,0)[r]{\strut{}$20$}}%
      \csname LTb\endcsname
      \put(3996,1088){\makebox(0,0)[r]{\strut{}$30$}}%
      \csname LTb\endcsname
      \put(3996,1326){\makebox(0,0)[r]{\strut{}$40$}}%
      \csname LTb\endcsname
      \put(3996,1564){\makebox(0,0)[r]{\strut{}$50$}}%
      \csname LTb\endcsname
      \put(4176,154){\makebox(0,0){\strut{}$1$}}%
      \csname LTb\endcsname
      \put(4776,154){\makebox(0,0){\strut{}$10$}}%
      \csname LTb\endcsname
      \put(5375,154){\makebox(0,0){\strut{}$100$}}%
      \put(581,1778){\makebox(0,0)[l]{\strut{}strong}}%
      \put(2379,1778){\makebox(0,0)[l]{\strut{}weak 18MB}}%
      \put(4382,1778){\makebox(0,0)[l]{\strut{}weak 180MB}}%
      \put(-448,445){\rotatebox{90}{\makebox(0,0)[l]{\strut{}GFLOPS/core}}}%
    }%
    \gplgaddtomacro\gplfronttext{%
      \csname LTb\endcsname
      \put(4914,-176){\makebox(0,0){\strut{}Number of Nodes}}%
      \csname LTb\endcsname
      \put(5333,764){\makebox(0,0)[r]{\strut{}CTF-def}}%
      \csname LTb\endcsname
      \put(5333,544){\makebox(0,0)[r]{\strut{}CTF-na}}%
    }%
    \gplbacktext
    \put(0,0){\includegraphics{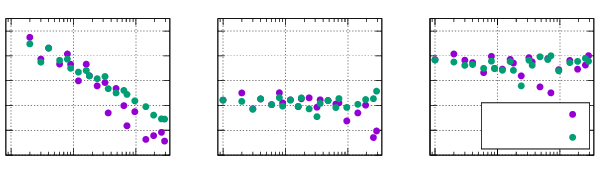}}%
    \gplfronttext
  \end{picture}%
\endgroup